\date{}
\documentclass[showpacs]{revtex4}
\usepackage{graphicx,psfrag,amsmath,amssymb,amsfonts,bbm,latexsym,color,dcolumn,
epsf}
\begin{document}
\title{Casimir energies with finite-width mirrors}
\author{C. D. Fosco$^a$ }
\author{F. C. Lombardo$^b$ }
\author{F. D. Mazzitelli$^b$ }
\affiliation{$^a$Centro At\'omico Bariloche and Instituto Balseiro,
 Comisi\'on Nacional de Energ\'\i a At\'omica, \\
R8402AGP Bariloche, Argentina}
\affiliation{$^b$Departamento de
F\'\i sica {\it Juan Jos\'e Giambiagi}, FCEyN UBA, Facultad de
Ciencias Exactas y Naturales, Ciudad Universitaria, Pabell\' on I,
1428 Buenos Aires, Argentina.}
\date{today}
\begin{abstract}
\noindent We use a functional approach to the Casimir effect in
order to evaluate the exact vacuum energy for a real scalar field
in $d+1$ dimensions, in the presence of  backgrounds that, in a
particular limit, impose Dirichlet boundary conditions on one or
two parallel surfaces.  Outside of that limit, the backgrounds are
described by a nonlocal effective action and may be thought of as
modelling finite-width mirrors with frequency-dependent
transmission and reflection coefficients. We obtain formal
expressions for the Casimir energy in general backgrounds, and
provide new explicit results in some particular cases.
\end{abstract}
\maketitle
\section{Introduction}\label{sec:intro}
The last years have seen  a renewed interest in the Casimir effect ~\cite{rev}.  The
main reason for this has been the design and successful implementation of
precision experiments~\cite{exp}. The magnitudes involved in those
experiments posed an important challenge to the first theoretical
calculations, usually based on simplified theoretical models. Indeed, in
those models, the materials were usually regarded as {\em perfect
mirrors\/}, imposing strict boundary conditions on the quantum fields.

 In order to explain the experimental results, however, it is important to
use more accurate models, including the corrections due to, for example,
rugosity, and to finite temperature and conductivity. In this
respect, there is a running controversy about how to model theoretically
the finite conductivity of the material, and also about its influence on
the Casimir forces~\cite{cont}.  The computation of the self-energies due
to quantum fluctuations is even more controversial: while the Casimir
forces between two objects become finite and independent on the properties
of the materials in the limit of perfect conductivity, the situation for
the self-energy is, by far, not so clear. The reason may be traced back to
the fact that the local energy density diverges close to an idealized
boundary~\cite{candelas}, and it is therefore necessary to introduce {\em
surface\/} counterterms in order to get finite results for the
self-energy~\cite{milton}.  Since the self-energy is relevant, for
instance, to analyze physical problems like gravitational effects produced
by the vacuum energy, a careful analysis of the sources of divergences in
the stress tensor as well as in the total energy is in order.

Motivated by those problems, in this paper we are concerned with the
calculation of the vacuum energy distortion that results from the presence
of some particular backgrounds, which are meant to represent more realistic boundaries,
for a real scalar field theory in $d+1$ spacetime dimensions. Following~\cite{Graham},
we will choose backgrounds which may be used to reproduce exact Dirichlet boundary conditions on flat
surfaces, when a particular limit for the parameters that define the
background is taken (see also~\cite{milton}).  As in~\cite{Graham}, we also
introduce two properties in the descriptions of the plates: finite widths
and finite strengths.  The former is represented by a smooth function
$\sigma_\epsilon$, depending on a single variable (the normal coordinate to
the plate), which has a characteristic width $2\epsilon$. Regarding the
strength of the coupling between the walls and the fluctuating field,  we
extend the kind of background considered in~\cite{Graham}, by allowing for
a frequency and (parallel) momentum dependent form for a function
$\lambda$, which determines the strength of that coupling.
Any non-trivial dependence on those variables shall amount, in coordinate space,
to a nonlocal form for the action that describes the interaction between field and plates.
The main new contribution of this paper is the derivation of
expressions for the Casimir energy, taking into account nonlocal effective actions that
model finite-width mirrors with frequency dependent reflection and
transmission coefficients, using the functional formalism as a tool. We do that for general cases, obtaining formal expressions, as well as for particular examples, where we derive more concrete results.

We will also be able to consider  smooth backgrounds, in order to test the
dependence of the Casimir energy with the `sharpness' of the
boundaries. As already stressed, these generalizations are of
interest for the calculation of the Casimir forces in realistic
situations, and also from a formal point of view, since many of
the divergences that appear in the calculations of the zero point
self-energies are related to the introduction of sharp interfaces
and/or ideal boundary conditions. In this way, with finite-widths
and frequency-momentum dependent strengths, one can simultaneously
control the two sources of UV divergences that pop-up during the
calculation of vacuum energies. This will also allow us to avoid
the unnecessary introduction of ill-defined quantities at the
intermediate steps, in the course of calculating physical
observables.

The calculation of the Casimir energy in the presence of realistic
mirrors is a subject that have been considered in a large number
of previous papers. In particular, within the Lifshitz approach it is,
in principle, possible to compute the zero point energies for slabs
of arbitrary width and arbitrary electromagnetic properties,
bounded by flat interfaces~\cite{lif}. In our approach, we will be
able to consider smooth surfaces, including some which approximate slabs
(to an arbitrary degree). Moreover, the functional method is particularly
well adapted to consider the effects on the zero point energy of the degrees of
freedom inside the material. Indeed, the nonlocal effective action
may be thought of as resulting precisely from the integration of those
degrees of freedom. Thus our initial nonlocal effective action can
be considered as a toy model for the interaction of the would-be
electromagnetic field (here the quantum scalar field) and the
charges confined in the material. In order to illustrate the
method, in the present paper we will consider a particular class
of nonlocal effective actions for a scalar field, leaving the
generalization to the electromagnetic case and the derivation of
the effective action from first principles for a future work.
Nonlocal effective actions have also been briefly considered by
other authors, see for instance~\cite{2nonlocal}.

>From a more technical point of view, we shall use an extended
version of a previously used functional approach \cite{funct} to
the Casimir effect with perfect mirrors, in order to cope with the
more general situations considered here.

The organization of this paper is as follows: in Section
\ref{sec:method}, we introduce the general method used to
calculate Casimir energies. We obtain formal expressions for the
total vacuum energy for one and two mirrors. We then present
explicit results: in Section \ref{sec:single}, for the change in
the vacuum energy under the introduction of a single mirror, to
consider afterwards, in Section \ref{sec:double}, the case of two
mirrors. For the single mirror case, we compute the Casimir energy
associated to smooth and piecewise constant backgrounds. We
discuss the dependence of the energy with the width $\epsilon$,
the cutoff frequency, and the sharpness of the boundary.  For the
case of two mirrors, we present a general expression, suitable for
numerical evaluation, and derive some of its general properties.
The expression becomes much more simple for the particular
situation in which the width of the mirrors is much smaller than
the distance between them, but not necessarily smaller than the
inverse of the strength of the coupling $\lambda^{-1}$. In Section
\ref{sec:concl} we summarize our conclusions.
\section{The method}\label{sec:method}
In this Section we set up the problem, and derive the general expressions
subsequently used for the calculation of vacuum energies in some specific
cases.

We shall consider either one or two identical, flat, parallel and finite
width ($\sim \, 2\epsilon$) mirrors in $d$ spatial dimensions. A coordinate
system has been chosen such that their `centers' correspond to $x_d = 0$
for the case of a single mirror, and to $x_d=0$ and $x_d=a$, for the case
of two parallel mirrors.  In both cases, the system will be described by an
Euclidean action:
\begin{equation}\label{eq:defs}
S(\varphi) \;=\; S_0(\varphi) \,+\, S_I(\varphi) \;,
\end{equation}
where $S_0$ defines the free theory:
\begin{equation}\label{eq:defs0}
S_0(\varphi)\;=\; \frac{1}{2} \, \int d^{d+1}x \; \partial_\mu \varphi
\partial_\mu \varphi \;,
\end{equation}
while $S_I(\varphi)$ is a term, quadratic in the field, that
contains the interaction with the walls. Introducing a number $N=1,2$
depending on whether one has one or two walls, respectively, we have:
\begin{equation}
S_I=\sum_{\alpha=1}^N S_I^{(\alpha)} \;,
\end{equation}
where:
\begin{equation}\label{eq:defsia}
  S_I^{(\alpha)} (\varphi)=\frac{1}{2}\int dx_0 \int dx'_0 \int d^{d -
1}x_\parallel \int d^{d-1}x'_\parallel \int dx_d \; \varphi(x_0,{\mathbf
x}_\parallel, x_d) \, \lambda(x_0-x_0'; {\mathbf x}_\parallel - {\mathbf
x}_\parallel')\sigma_\epsilon(x_d - a_\alpha) \,\varphi(x'_0,{\mathbf
x}_\parallel ', x_d)\;,
\end{equation}
with $a_1\equiv 0$ and $a_2\equiv a$. ${\mathbf x}_\parallel$ denotes the
$d-1$ coordinates parallel to the mirror: $x_1,x_2,\ldots, x_{d-1}$. We
regard the interaction term as an effective action coming from the
integration of the degrees of freedom confined to the walls that interact
with the scalar field $\varphi (x)$. On general grounds, the quadratic part
of that effective action should be of the form:
\begin{equation}
{\cal S}_{\rm eff} \,=\,\frac{1}{2} \int d^{d+1}x\,\int d^{d+1}x'\,\,
\varphi(x) \Gamma_2 (x ; x') \varphi(x')\,\,.
\end{equation}
Taking into account translation invariance in $x_0$ and ${\mathbf
x}_\parallel$, the dependence of the kernel $\Gamma_2$ on the coordinates
must be of the form:
\begin{equation}
\Gamma_2 (x ; x') = \Gamma_2 (x_d, x'_d, x_0 - x'_0, {\mathbf
x}_\parallel - {\mathbf x'}_\parallel)\,.
\end{equation}
If, in addition, we assume locality in the perpendicular
coordinate $x_d$, we arrive at (\ref{eq:defsia}), where the
functions $\lambda$ and $\sigma_\epsilon$ encode the structure of
the walls, such as their reflection and transmission coefficients
and their widths. We will assume, without any lose of generality,
that $\lambda$ is an even function (any odd part would cancel away
in the action). Therefore its Fourier transform,
${\tilde\lambda}$, is real. Besides, we will also assume that
${\tilde\lambda}$ is strictly positive, so that the interaction
with the walls is repulsive at all frequencies. Note that the
nonlocality in configuration space will produce reflection and
transmission coefficients dependent on the frequency $\omega$, as
well as and on the wave vector along the parallel plane, ${\mathbf
k}_\parallel$.

On the other hand, $\sigma_\epsilon$ is an even, strictly positive and
continuous function, approximately constant on a region of size $\sim 2
\epsilon$ around $0$. There is no big loose of generality by assuming
strict positivity for this function.  Indeed, assuming that one wanted to
consider, for example, a function whose support is the interval
$[-\epsilon,\epsilon]$, we could approximate it by a strictly positive one
that vanished, arbitrarily fast, outside that interval (see
Section~\ref{sec:single} where this case is dealt with in some detail).
Since $\sigma_\epsilon$ appears multiplied by $\lambda$, we may impose the
condition:
\begin{equation}
\int_{-\infty}^{+\infty}dx_d \, \sigma_\epsilon (x_d) = 1 \;,
\end{equation}
without restricting the actual interaction at all. Rather, it is a way of
disentangling  the `shape' (attributed to $\sigma_\epsilon$) from the strength
(carried by $\lambda$) of the interaction.

In our conventions, the Euclidean coordinates shall be denoted as $x_\mu$,
\mbox{$\mu=0,1,\ldots, d$}. The $x_d$ coordinate points along the normal
direction to the walls, while \mbox{${\mathbf x}_\parallel \equiv
(x_1,\ldots, x_{d-1})$} are parallel to them. The $d$ spatial coordinates
are collectively denoted by ${\mathbf x}$.

We now consider the vacuum energy cost, of distorting the vacuum
by the introduction of the mirrors. This quantity may be written
as follows:
\begin{equation}\label{eq:e0}
  E_0 \;=\; - \, \lim_{T \to \infty} \, \frac{1}{T} \,
\ln \big(\frac{\mathcal Z}{{\mathcal Z}_0} \big)
\end{equation}
where
\begin{equation}\label{eq:defzz0}
  {\mathcal Z}= \int {\mathcal D}\varphi \, e^{-S(\varphi)} \;\;,\;\;\;
  {\mathcal Z}_0 = \int {\mathcal D}\varphi \, e^{-S_0(\varphi)} \;,
\end{equation}
and $T$ is the extension of the (imaginary) time interval.

In more than one spatial dimension, $E_0$ is proportional to the `area'
$L^{d-1}$ of the mirrors (assumed to be $(d-1)$-dimensional squares of side
$L$). Since $L\to \infty$, it is convenient to introduce the energy density
${\mathcal E}_0$, such that
\begin{equation}\label{eq:defz}
{\mathcal E}_0 \;\equiv\; \lim_{T,L\to \infty} \frac{1}{L^{d-1} T} \,
\ln \big(\frac{\mathcal Z}{{\mathcal Z}_0}\big) \;.
\end{equation}

Besides, in the case of two mirrors, one is usually interested not in
${\mathcal E}_0$, but rather in a subtracted quantity, $\tilde{\mathcal
E}_0$, defined as the difference:
\begin{equation}
\tilde{\mathcal E}_0 \;\equiv\; {\mathcal E}_0 \,-\,{\mathcal E}_0 (\infty)
\end{equation}
where ${\mathcal E}_0 (\infty)$ denotes the surface energy density when the
mirrors are separated by an infinite distance.

$\tilde{\mathcal E}_0$ is finite even if ideal mirrors (i.e.,
imposing Dirichlet boundary conditions) were considered. The
reason is that the self-energies of the mirrors, being translation
invariant, are cancelled when subtracting the $a \to \infty$
contribution. This, however, would involve a regulatization in
order to avoid the having to deal with the difference between two
ill-defined (divergent) quantities. In our case, the self-energies
are finite due to the presence of a physical regularization
mechanism. Therefore, the subtraction of the self-energies is a
well-defined step, without having to invoke any additional
regulator.

To proceed, we note that a quite natural extension of the functional
approach followed in previous works \cite{funct} can be implemented here.
Indeed, we
can introduce an auxiliary field, $\xi_\alpha$, in order to rewrite the
exponential factor $S_I^{(\alpha)}(\varphi)$:
\begin{eqnarray}\label{eq:auxsingle}
  e^{-S_I^{(\alpha)}(\varphi)} &=&\frac{1}{\mathcal N} \,
  \int {\mathcal D}\xi_\alpha \,  e^{-\frac{1}{2} \int d^{d+1}x \int d^{d+1}x'
    \xi_\alpha(x) \lambda^{-1}(x_0-x_0'; {\mathbf x}_\parallel - {\mathbf
x}_\parallel ')
\sigma_\epsilon(x_d -a_\alpha)
 \delta (x_d- x_d') \xi_\alpha(x')} \nonumber\\
  &\times& e^{i \int d^{d+1}x \, \xi_\alpha(x) \, \sigma_\epsilon(x_d -
a_\alpha) \,
\varphi(x)} \;,
\end{eqnarray}
where the factor ${\mathcal N}$, independent of $\alpha$, is given by:
\begin{equation}
  {\mathcal N} \,=\, \int {\mathcal D}\xi_\alpha \, e^{-\frac{1}{2} \int
d^{d+1}x \int d^{d+1}x' \xi_\alpha(x) \lambda^{-1}(x_0-x_0'; {\mathbf
x}_\parallel - {\mathbf x}_\parallel ') \sigma_\epsilon(x_d) \delta (x_d-
x_d') \xi_\alpha(x')} \;,
\end{equation}
and we have used $\lambda^{-1}(x_0-x_0'; {\mathbf x}_\parallel - {\mathbf
x}_\parallel ')$ as a notation for the inverse kernel associated to
$\lambda (x_0-x_0'; {\mathbf x}_\parallel - {\mathbf x}_\parallel ')$,
i.e.,
\begin{eqnarray}
\int_{x_0'', {\mathbf x}_\parallel ''} \lambda(x_0-x_0''; {\mathbf x}_\parallel
- {\mathbf x}_\parallel '') \lambda^{-1}(x_0''-x_0'; {\mathbf x}_\parallel '' -
{\mathbf x}_\parallel ') &=&  \int_{x_0'', {\mathbf x}_\parallel ''}
\lambda^{-1} (x_0-x_0''; {\mathbf x}_\parallel - {\mathbf x}_\parallel '')
\lambda(x_0''-x_0'; {\mathbf x}_\parallel '' - {\mathbf x}_\parallel ')
\nonumber\\
&=& \delta(x_0-x_0') \delta ({\mathbf x}_\parallel
- {\mathbf x}_\parallel ') \;.
\end{eqnarray}

An important difference with the case of ideal, zero-width mirrors, is in
that the auxiliary fields now live in $d+1$ dimensions, rather than on the
$d$-dimensional submanifold $x_d=0$. The present case reduces to the ideal
situation if $\sigma_\epsilon$ is replaced by a $\delta$ function (of which
it may be regarded as an approximation when $\epsilon$ is finite). In other
words, there is a `dimensional reduction' in the auxiliary fields when
\mbox{$\epsilon \to 0$}. Particular profiles for the function
$\sigma_\epsilon$ are introduced in Sections~\ref{sec:single}
and~\ref{sec:double}, obtaining for them explicit results.

We now use (\ref{eq:auxsingle}) to rewrite ${\mathcal Z}$ and afterwards
integrate out the scalar field $\varphi$, to obtain:
\begin{equation}\label{eq:zaux1}
 {\mathcal Z} \;=\; {\mathcal Z}_0 \; \frac{1}{({\mathcal N})^N} \,
  \int \; \prod_{\alpha=1}^N {\mathcal D}\xi_\alpha \;
e^{-\frac{1}{2} \int d^{d+1}x \int d^{d+1}x' \sum_{\alpha, \beta = 1}^N \;
\xi_\alpha (x) \Omega_{\alpha \beta} (x,x') \xi_\beta (x') } \;,
\end{equation}
with a matrix kernel whose elements $\Omega_{\alpha\beta}$ are
defined by:
\begin{eqnarray}
\Omega_{\alpha\beta}(x,x') \; &=& \;  \delta_{\alpha\beta} \,
\lambda^{-1} (x_0-x_0'; {\mathbf x}_\parallel - {\mathbf x}_\parallel ')
\sigma_\epsilon(x_d-a_\alpha)
\delta (x_d - x_d') \nonumber \\
&+& \sigma_\epsilon(x_d-a_\alpha) \Delta(x_0,{\mathbf x}_\parallel, x_d ;
x_0',{\mathbf x'}_\parallel, x_d' ) \sigma_\epsilon(x_d'-a_\beta) \;,
\end{eqnarray}
where $\Delta$ is the free scalar-field propagator:
\begin{equation}
  \Delta(x,x') = \Delta(x-x')= \int \frac{d^{d+1}k}{(2\pi)^{d+1}}
e^{i k \cdot (x-x')} \frac{1}{k^2} \;.
\end{equation}

We then make, for each auxiliary field $\xi_\alpha$, the redefinition:
\begin{equation}
  \xi_\alpha (x_0,{\mathbf x}_\parallel , x_d) \to  \int_{x_0', {\mathbf
x}_\parallel '} \,
\lambda^{\frac{1}{2}}(x_0-x_0'; {\mathbf x}_\parallel - {\mathbf x}_\parallel ')
\,
\xi_\alpha (x_0',{\mathbf x}_\parallel ', x_d - a_\alpha) \;,
\end{equation}
both in the explicit integral over $\xi$ in (\ref{eq:zaux1}) and in the
implicit one in the definition of ${\mathcal N}$. We have assumed that the
square-root kernel is well-defined, what is consistent with the assumptions
about its properties.  This redefinition produces a non-trivial
($\lambda$-dependent) Jacobian for each factor ${\mathcal D}\xi_\alpha$. An
identical Jacobian does, however, appear also in each denominator
${\mathcal N}$, and therefore they cancel each other.  Besides, after the
redefinition, the ${\mathcal N}$ factor changes: ${\mathcal N}\to
\widetilde{\mathcal N}$, which is independent of $\lambda$.  Thus,
after a trivial shift in the coordinates,
\begin{equation}\label{eq:zaux2}
  {\mathcal Z} \;=\; {\mathcal Z}_0 \;\frac{1}{(\widetilde{\mathcal N})^N} \;
  \int \prod_{\alpha=1}^N {\mathcal D}\xi_\alpha \,
e^{-\frac{1}{2} \int d^{d+1}x \int d^{d+1}x' \sum_{\alpha,\beta} \xi_\alpha (x)
    \tilde{\Omega}_{\alpha\beta}(x,x') \xi_\beta (x') } \;,
\end{equation}
where:
\begin{eqnarray}
\widetilde{\Omega}_{\alpha\beta}(x,x') &=& \delta_{\alpha\beta} \;
\delta^{(d+1)}(x-x') \sigma_\epsilon(x_d)
\nonumber\\
&+&\sigma_\epsilon(x_d) \int_{x_0'',x_0''', {\mathbf x}_\parallel '', {\mathbf
x}_\parallel '''} \lambda^{\frac{1}{2}}(x_0-x_0''; {\mathbf x}_\parallel -
{\mathbf x}_\parallel '')
\Delta(x_0'',{\mathbf x}_\parallel '', x_d + a_\alpha ;x_0''',{\mathbf
x}_\parallel ''', x_d + a_\beta ) \nonumber\\
&\times& \lambda^{\frac{1}{2}}(x_0'''- x_0'; {\mathbf x}_\parallel '''- {\mathbf
x}_\parallel ') \, \sigma_\epsilon(x_d') \;.
\end{eqnarray}

We then proceed to perform yet another redefinition of the auxiliary fields, now
involving a diffeomorphism in the $x_d$ coordinate. We introduce the
one-to-one mapping $x_d \to z$ that results from the differential equation:
\begin{equation}\label{eq:defzxd}
\frac{dz}{dx_d} \;=\; \sigma_\epsilon(x_d) \;,
\end{equation}
where the assumed positivity of $\sigma_\epsilon$ comes in handy to ensure
that the change of variables is non-singular.  Imposing the condition
$z(0)=0$, the solution to the previous equation is unique. We also note
that, as a consequence of the change of variables,
\begin{equation}
\delta(x_d - x_d') \,=\, \sigma_\epsilon(x_d) \, \delta(z-z') \;\;,\;\;\;
dx_d \, \sigma_\epsilon(x_d) \,=\, dz \,.
\end{equation}
Equation (\ref{eq:defzxd}), together with the condition $z(0) = 0$ and the
normalization condition for $\sigma_\epsilon$ imply that the range of $z$
is (regardless of the particular profile used for $\sigma_\epsilon$) the
finite interval $[-\frac{1}{2},\frac{1}{2}]$.

Keeping the same notation for the auxiliary fields when written in terms of
the new variables, we then have, in a simplified notation:
\begin{equation}\label{eq:zaux3}
{\mathcal Z} \;=\; {\mathcal Z}_0 \;
\int \prod_{\alpha=1}^N {\mathcal D}\xi_\alpha \,
e^{-\frac{1}{2} \int_{x_0,{\mathbf x_\parallel}, z;
x'_0,{\mathbf x'_\parallel}, z'} \sum_{\alpha,\beta} \xi_\alpha (x_0,{\mathbf
x_\parallel}, z)
{\mathcal K}_{\alpha\beta}(x_0,{\mathbf x}_\parallel,z \,;\,
 x_0',{\mathbf x'_\parallel},z') \xi_\beta(x_0',{\mathbf x'_\parallel}, z') }
\;,
\end{equation}
with
\begin{eqnarray}
&&{\mathcal K}_{\alpha\beta}(x_0,{\mathbf x}_\parallel,z; x_0',{\mathbf
x'_\parallel},z')\; = \;
\delta_{\alpha\beta} \, \delta(x_0-x'_0) \delta^{(d-1)}({\mathbf
x_\parallel}-{\mathbf x'_\parallel})
\delta(z-z') \nonumber \\
&+& \int_{x_0'',x_0''', {\mathbf x}_\parallel '', {\mathbf x}_\parallel '''}
\lambda^{\frac{1}{2}}(x_0-x_0''; {\mathbf x}_\parallel - {\mathbf x}_\parallel
'')
\Delta(x_0'',{\mathbf x_\parallel},x_d(z) + a_\alpha ;\,
x_0''',{\mathbf x'_\parallel},x'_d(z') + a_\beta )
\lambda^{\frac{1}{2}}(x_0'''-x_0'; {\mathbf x}_\parallel '''- {\mathbf
x}_\parallel ') \;.
\end{eqnarray}
We note that, had any Jacobian arisen because of the last field
diffeomorphism, it would have, again, been cancelled against an equal
object coming from $\widetilde{\mathcal N}$. Furthermore, since after the
redefinition $\widetilde{\mathcal N}$ became independent of $\lambda$ and
$\sigma_\epsilon$, that factor has been dropped.

Then we Fourier transform all the spacetime coordinates for which there is
translation invariance, namely, $x_0,x_\parallel$. The transformed
operator has the following form:
\begin{equation}\label{eq:defk}
\widetilde{\mathcal K}_{\alpha\beta} (z;z')\;=\; \delta_{\alpha\beta} \,
\delta(z-z')
\,+\, {\tilde \lambda}(\omega, {\mathbf k}_\parallel) \,
D_{\alpha\beta}(z;z')\;,
\end{equation}
where
\begin{equation}
{\tilde \lambda}(\omega, {\mathbf k}_\parallel) =
\int_{-\infty}^{+\infty} dx_0 \, \int d^{d-1}x_\parallel\,\, e^{-i
\left(\omega x_0 + {\mathbf k}_\parallel {\mathbf x}_\parallel
\right)} \lambda(x_0, {\mathbf x}_\parallel)
\end{equation}
and
\begin{equation}
D_{\alpha\beta}(z;z') \equiv \frac{e^{- \kappa |x_d(z)- x_d(z') +
a_\alpha - a_\beta| }}{2 \kappa} \;,
\end{equation}
where \mbox{$\kappa \equiv \sqrt{\omega^2 + k_\parallel^2}$} (to simplify
the notation, we omit writing the dependence of $D_{\alpha\beta}(z;z')$ on
$k_\parallel$ and $\omega$ explicitly).

Then we have the following expression for ${\mathcal Z}$:
\begin{equation}\label{eq:zk}
{\mathcal Z} \;=\; {\mathcal Z}_0 \; \big( \det \widetilde{\mathcal K}
\big)^{-\frac{1}{2}} \;,
\end{equation}
and:
\begin{equation}
  {\mathcal E}_0 \;=\; \lim_{T, L \to \infty} \; \frac{1}{2 T L^{d-1}}  \,
{\rm Tr} \big(\ln \widetilde{\mathcal K} \big) \;,
\end{equation}
where `${\rm Tr}$' denotes  trace over all the variables (including
$\omega$ and $k_\parallel$). Introducing the symbol `$\widetilde{\rm Tr}$'
for the (reduced) trace over the Hilbert space of functions depending on $z
\in [-\frac{1}{2}, \frac{1}{2}]$, we may write a more explicit formula,
where the trace over the variables for which the kernel is
translation-invariant is explicit:
\begin{equation}
{\mathcal E}_0 \;=\; \int \frac{d\omega}{2\pi}
\int \frac{d^{d-1}k_\parallel}{(2\pi)^{d-1}}\;
\widetilde{\rm Tr} \big(\ln \widetilde{\mathcal K} \big) \;.
\end{equation}
>From the last expression one can, in principle, extract the vacuum energy
relevant to each case. However, since there are important differences
between them, we present now separate calculations of ${\mathcal E}_0$,
corresponding to two different physical situations: for the case of one
mirror, and for a two-mirror system.
\subsection{One mirror} This case amounts to a single index
$\alpha=1$; thus we suppress the $\alpha$, $\beta$ indices, and
consider just the kernel:
\begin{equation}\label{eq:defk1}
\widetilde{\mathcal K} (z;z')\;=\; \delta(z-z')
\,+\, {\tilde \lambda}(\omega, {\mathbf k}_\parallel) D(z,z')\;,
\end{equation}
with
\begin{equation}
D(z;z') \,\equiv\,\frac{e^{- \kappa |x_d(z)- x_d(z')|}}{2 \kappa} \;.
\end{equation}
We shall calculate the trace of the logarithm of $\widetilde{\mathcal K}$,
by finding the eigenvalues of $\widetilde{\mathcal K}$.
The equation for $\psi_\alpha(z)$, the
eigenfunction associated to the eigenvalue $\alpha$, may be written as
follows:
\begin{equation}\label{eq:eigen1}
  \int_{-\frac{1}{2}}^{+\frac{1}{2}} dz' \, {\tilde \lambda}(\omega, {\mathbf
k}_\parallel) \,
 D(z;z') \, \psi_\alpha(z') \;=\; (\alpha - 1)
  \psi_\alpha(z) \;.
\end{equation}
This can be converted into a differential equation, taking into account the
fact that \mbox{$D(z;z')$} is obtained by performing a Fourier
transformation plus a change of variables in the Green's function for the
real scalar field. Indeed, starting from:
\begin{equation}
\big(-\frac{\partial^2}{\partial x_d^2} \,+\, \kappa^2 \big)
\;\tilde{\mathcal G}_0(x_d,x_d')\;=\; \delta(x_d-x_d') \;,
\end{equation}
where $\tilde{\mathcal G}_0(x_d,x_d')$ is  the Fourier transformed of the
free propagator:
\begin{equation}\label{eq:freef}
\tilde{\mathcal G}_0(x_d,x_d') \;\equiv\;
\frac{e^{- \kappa |x_d- x_d'|}}{2 \kappa} \;,
\end{equation}
we obtain, by changing variables:
\begin{equation}\label{eq:greendef}
\big[-{\tilde\sigma}_\epsilon (z) \frac{\partial}{\partial z} ( {\tilde
\sigma}_\epsilon (z) \frac{\partial}{\partial z} ) \,+\, \kappa^2 \big]
 D(z;z') \;=\; {\tilde\sigma}_\epsilon (z)\, \delta(z-z') \;,
\end{equation}
where
\begin{equation}
{\tilde\sigma}_\epsilon(z) \equiv \sigma_\epsilon(x_d(z)) \;.
\end{equation}

We see that, by acting with the differential operator that appears
on the left hand side of (\ref{eq:greendef}) on both sides of (\ref{eq:eigen1}),
it becomes a differential equation with the structure:
\begin{equation}\label{eq:sl1}
L[\psi_\alpha] (z) \;=\; \xi_\alpha \,\psi_\alpha (z) \;,
\end{equation}
where $L$ denotes a linear differential operator of the
Sturm-Liouville type:
\begin{equation}
L[f](z) \;=\; - \frac{d}{dz} \big[ p(z) \frac{df(z)}{dz} \big] \,+\,
q(z) \, f(z) \;,
\end{equation}
with
\begin{equation}
p(z) \equiv \tilde\sigma_\epsilon(z) \;\;,\;\;\; q(z) \equiv
\frac{\kappa^2}{\tilde\sigma_\epsilon}
\;,
\end{equation}
and the eigenvalue of $L$,  $\xi_\alpha$, determined $\alpha$
through the relation $\xi_\alpha = \frac{{\tilde
\lambda}}{\alpha-1}$.
The eigensystem corresponding to the operator whose determinant we need has
the coefficient functions $p$ and $q$, which are in turn determined by the
properties of the mirror.

An important fact to note is that the space of functions to be used for the
calculation of the eigenvalues of $L$ corresponds to functions that vanish
at $z = \pm \frac{1}{2}$. Indeed, the ${\widetilde K}$ operator is
self-adjoint under the $L^2({\mathbb R})$ scalar product defined by:
\begin{equation}
(f,g) \,\equiv\, \int_{-\infty}^{+\infty} dx_d \, f^*(x_d) g(x_d)
\end{equation}
for any pair of square integrable functions: $f$, $g$. This
scalar product becomes, when written in terms of the new variable $z$:
\begin{equation}\label{eq:defsp}
(f,g) \,\equiv\, \int_{-\frac{1}{2}}^{+\frac{1}{2}} \frac{dz}{{\tilde
\sigma}_\epsilon(z)} \, f^*(z) g(z) \;.
\end{equation}
In particular, this implies that normalizable functions in the interval
$[-\frac{1}{2},\frac{1}{2}]$ {\em must vanish\/} when $z=\pm 1/2$ (of
course, they should also decrease sufficiently fast for the integral to
converge). The reason is that, from the defining properties of $\sigma
_\epsilon$, it follows that ${\tilde \sigma}_\epsilon$ vanishes at least
linearly at those points.

\noindent It also implies that eigenfunctions of the Sturm-Liouville
operator corresponding to different eigenvalues shall be orthogonal, for
the scalar product (\ref{eq:defsp}), which includes the weight function
$1/{\tilde\sigma}_\epsilon$.

Thus, we conclude that, to evaluate the vacuum energy in the one mirror case,
this procedure has lead us to the expression
\begin{equation}\label{eq:onemir}
{\mathcal E}_0 \;=\; \frac{1}{2} \,
\int \frac{d\omega}{2\pi} \int \frac{d^{d-1}k_\parallel}{(2\pi)^{d-1}}
\sum_l\; \ln \big[\alpha_l(\omega, k_\parallel, \epsilon) \big]\;,
\end{equation}
where $l$ labels the eigenvalues $\alpha$. They are, in general, non trivial
functions of their arguments, and they depend on the particular
function $\sigma_\epsilon$ considered.

\subsection{Two mirrors} We shall now consider the situation of
two identical mirrors, located at a distance $a$ apart.  Thus, we
are concerned with the quantity:
\begin{equation}\label{eq:defe0}
\tilde{\mathcal E}_0(a) \;\equiv\; {\mathcal E}_0(a) \,-\, {\mathcal
E}_0(\infty)
\end{equation}
We have to consider now the $2 \times 2$ matrix kernel:
\begin{equation}
\big[{\widetilde{K}}_{\alpha\beta} (z;z')\big]\;=\;
 \left(
\begin{array}{cc}
\tilde{\mathcal K}(z;z') & {\mathcal P}(z;z') \\
{\mathcal Q}(z;z') & \tilde{\mathcal K}(z;z')
\end{array}
\right)
\end{equation}
where the two diagonal elements in the expression above coincide with the
(identically noted) kernel corresponding to the single-wall case, equation
(\ref{eq:defk1}). There appear also two new kernels ${\mathcal P}$ and
${\mathcal Q}$:
\begin{equation}
{\mathcal P}(z;z') \;=\;{\tilde\lambda}(\omega, {\mathbf k}_\parallel) \,
\tilde{\mathcal
G}_0\big(x_d(z)+0 ;
x_d(z')+a\big)\;=\;{\tilde\lambda}(\omega,{\mathbf k}_\parallel)
\, \frac{e^{- \kappa |x_d(z) - x_d(z') - a |}}{2 \kappa} \;,
\end{equation}
and ${\mathcal Q} \equiv {\mathcal P}|_{a \to -a}$.

One can then use the expression for the determinant of a
matrix in terms of a combination of its blocks,
\begin{equation}
\det \big[ K_{ij}\big] \;=\; \big(\det \tilde{\mathcal K}\big)^2 \, \det( 1 -
{\mathcal O} )
\end{equation}
where
\begin{equation}
{\mathcal O} \;=\; \tilde{\mathcal K}^{-1} \, {\mathcal P} \, \tilde{\mathcal
K}^{-1}
{\mathcal Q} \;.
\end{equation}
Then we note that, since $\det \tilde{\mathcal K}$ is independent of $a$,
the Casimir energy density, obtained by subtracting the contribution with the
plates at an infinite distance, becomes:
\begin{equation}\label{eq:tpe}
\tilde{\mathcal E}_0(a) \,=\, \lim_{T,L\to \infty} \,
\frac{1}{2 L^{d-1} T} \, {\rm Tr} \ln ( 1 - {\mathcal O} ) \;.
\end{equation}
It is important to note that, in the last equation, the $a$-independent
contribution, corresponding to the self-energies of the plates, have been
completely subtracted. Note that that contribution yields exactly twice the
vacuum energy for one plate. Of course, had $\lambda$ been assumed to be a
constant, those self energies would have been, generally, divergent.

Thus the UV behaviour of (\ref{eq:tpe}) is  milder that in other
approaches, where those contributions have to be dealt with in order to
give sense to the Casimir energy.  Here, as we shall see in the examples,
it is already finite. Assuming that
$\vartheta_l(\omega,k_\parallel,\epsilon, a)$ are the eigenvalues of
${\mathcal O}$:
\begin{equation}\label{eq:casimir}
\tilde{\mathcal E}_0(a) \,=\, \frac{1}{2} \, \int \frac{d\omega}{2\pi}
\int \frac{d^{d-1}k_\parallel}{(2\pi)^{d-1}}
\sum_l \, \ln \big[ 1 - \vartheta_l(\omega,k_\parallel, \epsilon, a)\big] \;,
\end{equation}
the expression that we will use as a starting point for the calculation of
the Casimir energy density in concrete examples.

The finiteness of (\ref{eq:casimir}) will follow from the fact that, as a
function of the frequency and momenta, ${\mathcal O}$ is bounded by
the exponentially decreasing factors of those variables, carried by
${\mathcal P}$ and ${\mathcal Q}$.
\section{Results for the total vacuum energy in the single-mirror
case}\label{sec:single}
\subsection{Smooth $\sigma_\epsilon$}
We first assume a particular form for the function $\sigma_\epsilon$. A
very convenient choice, from the calculational point of view is:
\begin{equation}\label{eq:defsigma}
\sigma_\epsilon (x_d) \;=\; \frac{1}{2\epsilon} \,
{\rm sech}^2(\frac{x_d}{\epsilon}) \;,
\end{equation}
where the $\frac{1}{2\epsilon}$ factor has been introduced in
order to satisfy \mbox{$\int_{-\infty}^{+\infty} dx_d
\sigma_\epsilon(x_d) = 1$}, the normalization condition
corresponding to an approximant of the $\delta$-function. Of
course, the wall is then localized around $x_d=0$, with a width
$\sim 2\epsilon$.
 Besides, for this profile we may find the explicit relation between
$x_d$ and $z$:
\begin{eqnarray}
x_d(z) &=& \epsilon \; {\rm arctanh}(2 z) \nonumber\\
z &=& \frac{1}{2} \tanh(\frac{x_d}{\epsilon}) \;.
\end{eqnarray}
Note that $z \in [-\frac{1}{2} , \frac{1}{2}]$. We immediately find:
\begin{equation}
{\tilde\sigma}_\epsilon(z) \,=\, \frac{1}{2\epsilon} \big[1 - (2 z)^2\big] \;.
\end{equation}

Applying the differential operator on the lhs of (\ref{eq:greendef}) to both
sides of (\ref{eq:eigen1}), and then changing variables: $z \to u \equiv 2 z$,
we obtain a differential equation for the eigenfunctions:
\begin{equation}
  \frac{d}{du} \big[(1-u^2) \frac{d}{du}\psi_\alpha(u)  \big] \,+\,
\big[\frac{\epsilon {\tilde \lambda}(\omega, {\mathbf k}_\parallel)}{2
(\alpha
- 1)} - \frac{(\epsilon \kappa)^2}{1 - u^2} \big] \psi_\alpha(u) \;=\; 0
  \;.
\end{equation}
We recognize the associated Legendre equation, which has regular independent
solutions when:
\begin{eqnarray}
  \frac{\epsilon {\tilde\lambda}(\omega, {\mathbf k}_\parallel)}{2 (\alpha - 1)}
&=& l (l +1)
  \;,\;\;\;\; l = 0,1,\ldots  \nonumber\\
  \epsilon \kappa &=& m \;,\;\;\;\; m = 1,\ldots, l \;.
\end{eqnarray}
Note that the $m=0$ solutions have been discarded, since the last condition
does not lead to any non-trivial solution. The eigenvectors are then the
polynomials $P_l^m(2 z)$ which (since $m \neq 0$) vanish at $z = \pm
\frac{1}{2}$, as expected from our general analysis at the end of the
previous section.

>From the above, we conclude that the eigenvalues are:
\begin{equation}
\alpha \,=\, \alpha(l,\omega) \,=\, 1 \,+\, \frac{\epsilon
  {\tilde \lambda}(\omega, {\mathbf k}_\parallel)}{2 l (l+1)} \;,
\end{equation}
while, for each $l$, we have the constraints:
\begin{equation}
\epsilon \sqrt{\omega^2 + k_\parallel^2}\,=\, 1,\ldots, l \;,
\end{equation}
which restrict the allowed values of $k_\parallel \equiv |{\mathbf
  k_\parallel}|$ and $\omega$.
We may then use the explicit form of the eigenvalues to obtain:
\begin{eqnarray}\label{eq:1mirrorfin}
  {\mathcal E}_0 &=& \frac{1}{2} \,
  \sum_{l=1}^\infty \sum_{m=1}^l \, \int
  \frac{d\omega}{2\pi} \int \frac{d^{d-1}k_\parallel}{(2\pi)^{d-1}} \nonumber\\
  &\times& \delta\big[\epsilon \sqrt{\omega^2 + k_\parallel^2} - m \big]\;
  \ln \big[  1 \,+\, \frac{\epsilon {\tilde\lambda}(\omega,{\mathbf
k}_\parallel)}{2 l (l + 1)}\big]
  \;,
\end{eqnarray}
Note that we are writing a general expression as a function of the
dimension of space, $d$, but we are not meaning by that that a
dimensional regularization is used. Indeed, as we will see, this
expression will be regularized by the function $\tilde\lambda$.

In the case $d=1$, there is no integral over $k_\parallel$.
Furthermore, the integral over $\omega$ becomes trivial
because of the constraint, and thus we derive an expression for the vacuum
energy in terms of a double sum:
\begin{eqnarray}
    {\mathcal E}_0 = E_0 = \frac{1}{4\pi \epsilon} \sum_{l=1}^\infty
    \sum_{m=1}^l &\Big\{& \ln \big[ 1 + \frac{\epsilon
      \tilde\lambda(\frac{m}{\epsilon})}{2 l (l+1)}  \big]
      \nonumber\\
      &+& \ln \big[ 1 + \frac{\epsilon \tilde\lambda(-\frac{m}{\epsilon})}{2 l
        (l+1)} \big] \Big\} \;,
\end{eqnarray}
or,
\begin{equation}\label{eq:resd1}
    {\mathcal E}_0 = \frac{1}{4\pi\epsilon} \sum_{l=1}^\infty
    \sum_{m=1}^l \ln \big| 1 + \frac{\epsilon
    \tilde\lambda(\frac{m}{\epsilon})}{2 l (l + 1)} \big|^2 \;,
\end{equation}
since $\lambda(x_0-x_0')$ is assumed to be real (i.e., no
dissipative coupling to the mirror).

The result (\ref{eq:resd1}) shows an interesting relationship
between the large-$\omega$ behaviour of $\tilde\lambda$ and the
size of the mirror, $\epsilon$. Indeed, assume that
$\tilde\lambda$ becomes negligible small above some cutoff
frequency $\omega_c$, then we have the approximate expression:
\begin{equation}
  {\mathcal E}_0 \;\simeq\;\frac{1}{4\pi \epsilon} \,\sum_{l=1}^\infty \,
  \; \sum_{m=1}^{{\rm min} \{l,[\epsilon \omega_c] \}} \,
\ln \big| 1 + \frac{\epsilon
      \tilde\lambda(\frac{m}{\epsilon})}{2 l (l + 1)} \big|^2 \;.
\end{equation}
Note that the energy is zero if \mbox{$\epsilon < 1/\omega_c$}.
This may be understood from the fact that no modes could be
trapped inside the mirror, since it is transparent for wavelengths smaller
than $\epsilon$.

To proceed in the case $d > 1$, we will assume that
${\tilde\lambda}(\omega,{\mathbf
k}_\parallel)={\tilde\lambda}(\kappa)$. The motivation for this
assumption is that, with this choice, the reflection and
transmission coefficients will depend on the wave number $k_d$ in
the perpendicular direction (note that when rotated back to
Minkowski space, $\kappa$ becomes $k_d$). In this case, the
integrals in Eq.(\ref{eq:1mirrorfin}) can be easily computed. We
obtain
\begin{equation}
    {\mathcal E}_0  = \frac{\Omega_{d-1}} {2(2\pi\epsilon)^d} \sum_{l=1}^\infty
    \sum_{m=1}^l  m^{d-1}\;
    \ln \big[ 1 + \frac{\epsilon \tilde\lambda(\frac{m}{\epsilon})}{2 l (l+1)}
\big]
    \;,
\end{equation}
where $\Omega_{d-1}=...$ is the solid angle in $d-1$ dimensions.
If there is a cutoff $\omega_c$, as in the previous example, and
besides we have that for $\kappa < \omega_c$ it is constant:
$\tilde\lambda = \tilde \lambda_0$, then:
\begin{equation}\label{eq:aux11}
  {\mathcal E}_0 \,=\, \frac{\Omega_{d-1}}{2 (2\pi\epsilon)^d}
\sum_{l=1}^\infty
  \ln \big[ 1 + \frac{\epsilon \tilde\lambda_0}{2 l (l+1)}  \big] \,
  \big( \sum_{m=1}^{{\rm min} \{l,[\epsilon \omega_c] \}}  m^{d-1} \big)\;.
\end{equation}
Of course, we also have the same relationship between $\omega_c$
and $\epsilon$ as in the previous example with $d=1$,
since the defect is essentially one-dimensional.

Let us see that the introduction of the cutoff $\omega_c$ produces a finite result, in
any number of dimensions. To see this, we just consider the case of a $\tilde\lambda$ which is constant below the cutoff,
and zero above. Then we may rewrite (\ref{eq:aux11}) more explicitly, as follows:
\begin{eqnarray}\label{eq:aux12}
  {\mathcal E}_0 &=& \frac{\Omega_{d-1}}{2 (2\pi\epsilon)^d}
\sum_{l=1}^{[\epsilon\omega_c]}  \ln \big[ 1 + \frac{\epsilon \tilde\lambda_0}{2 l (l+1)}  \big] \,
  \big( \sum_{m=1}^l  m^{d-1} \big) \nonumber\\
&+& \frac{\Omega_{d-1}}{2 (2\pi\epsilon)^d} \big( \sum_{m=1}^{[\epsilon\omega_c]} m^{d-1} \big)
\sum_{l=[\epsilon\omega_c]+1}^\infty  \ln \big[ 1 + \frac{\epsilon \tilde\lambda_0}{2 l (l+1)}  \big]
\;.
\end{eqnarray}
Now, the first term on the rhs above is a finite sum, while the second one involves a convergent series, so the result is finite. Of course, the
same holds true for the case of a (continuous) $\tilde\lambda$ function which is not necessarily constant below the cutoff. Indeed, one may simple
bound that function by a constant, and then use the previous result.
\subsection{Piecewise constant profile}
We consider now another important profile for the function
$\sigma_\epsilon$. We assume that this function takes the
approximately constant value $1/(2 \epsilon)$ inside the interval
$[-\frac{1}{2},\frac{1}{2}]$, and vanishes very fast outside. The
function is then extremely smooth, except for small intervals
around $x_d =\pm \epsilon$ where all the variation is
concentrated. Of course, one can approximate a `square barrier'
function with this kind of profile, and this is the motivation for
considering it.

In the limit when that profile is approached, we are left with the
equation for the eigenvalues:
\begin{equation}
-\frac{d^2}{dz^2} \psi_\alpha(z) + (2 \epsilon \kappa)^2
\psi_\alpha(z) \;=\; \frac{2 \epsilon
\tilde{\lambda}(\omega,{\mathbf k}_\parallel)}{\alpha-1}
\psi_\alpha
\end{equation}
in the interval $(-\frac{1}{2},\frac{1}{2})$. With the boundary conditions
we find the eigenvalues of ${\tilde{\mathcal K}}$, which form a discrete set:
\begin{equation}\label{eq:eigensq}
\alpha \;=\; \alpha_l \,=\, 1 \,+\, \frac{2 \epsilon
{\tilde\lambda}(\omega,{\mathbf k}_\parallel)}{l^2 \pi^2 \,+\, (2
\epsilon \kappa)^2} \;, \;\;\;\; l \in {\mathbb N}\;.
\end{equation}
Thus, in $d$ spatial dimensions, the expression for ${\mathcal E}$ becomes:
\begin{equation}\label{eq:e0s}
{\mathcal E}_0 \;=\; \frac{1}{2} \; \int \frac{d\omega}{2\pi} \int
\frac{d^{d-1}k_\parallel}{(2\pi)^{d-1}} \, \sum_{l=1}^\infty \, \ln \left[
 1 \,+\, \frac{2 \epsilon {\tilde\lambda}(\omega,{\mathbf k}_\parallel)}{l^2
\pi^2 + (2
\epsilon)^2 (\omega^2 + k_\parallel^2)} \right] \;
\end{equation}
which differs from the one obtained for the other profile used for
$\sigma_\epsilon$.  It should be noted that the procedure above for the
square barrier profile relies upon the existence of a region where that
function is approximately constant. This would not make sense if we wanted
to consider the $\delta$-function ($\epsilon \to 0$) limit, which we shall
study (see next section) using the $({\rm sech})^2$ function instead.

We assume, as stated before, that $\tilde{\lambda}$
depends on $\omega$ and $k_\parallel$ trough the particular combination
\mbox{$\kappa = \sqrt{\omega^2 + k_\parallel^2}$}, i.e., $\tilde{\lambda} =
\tilde{\lambda}(\kappa)$. Then we may write a more explicit formula for ${\mathcal E}_0$:
\begin{equation}
{\mathcal E}_0 \;=\; \frac{1}{2^d \pi^{d/2} \Gamma(d/2)} \,
\int_0^\infty d\kappa \, \kappa^{d-1} \;  \sum_{l=1}^\infty \, \ln \left[
 1 \,+\, \frac{2 \epsilon {\tilde\lambda}(\kappa)}{l^2
\pi^2 + (2 \epsilon \kappa)^2} \right] \;.
\end{equation}
The series is convergent, and one can immediately find its large-$\kappa$ behaviour:
\begin{equation}
\sum_{l=1}^\infty \, \ln \left[
 1 \,+\, \frac{2 \epsilon {\tilde\lambda}(\kappa)}{l^2
\pi^2 + (2 \epsilon \kappa)^2} \right] \,\sim\, \frac{\tilde\lambda(\kappa)}{\kappa} \;.
 \end{equation}
>From the previous expression, we see that if $\tilde\lambda$ vanishes above a cutoff, the energy is finite. We see that it would also be finite if it vanished following an exponential law,
or even a power-law: $\tilde\lambda \sim 1/\kappa^\alpha$, with $\alpha > d - 1$.
\subsection{UV divergences}
We conclude this section by presenting a brief study of the UV behaviour (divergences) of the expressions
 that we used to calculate the energies in the one mirror case; i.e., of the self-energies,
when one approaches the constant-$\tilde\lambda$ case.

We will study here that limit starting form the self-energy for a finite-width mirror with a cutoff for $\kappa$, so that the energy is finite, and then take the relevant limit, which  corresponds to a constant $\tilde\lambda$ (that eventually tends to infinity), and a vanishing $\epsilon$.
The most transparent way to take these two limits is by letting first $\epsilon \to 0$, and then making $\tilde\lambda$ go to a constant. The advantage of using this order is that the first limit can be taken exactly. Indeed, when $\epsilon \to 0$,
we have:
\begin{equation}
\tilde{\mathcal K}(z;z') \,\to \, \delta(z-z') \,+\,
\frac{\tilde\lambda(\kappa)}{2\kappa}\;,
\end{equation}
or, in operatorial form,
\begin{equation}
\tilde{\mathcal K}\,\to \, (1\,+ \,\frac{\tilde\lambda(\kappa)}{2\kappa}) \rho
\, + \, \eta \;,
\end{equation}
where $\rho$ and $\eta$ are projection operators whose kernels are:
\begin{equation}
\rho(z;z') \;=\; 1 \;\;,\;\;\;\; \eta(z;z') \;=\; \delta(z-z') - \rho(z,z')
\;.
\end{equation}

Thus, the expression for the vacuum energy in this limit is
\begin{equation}
\big[{\mathcal E}_0\big]_{\epsilon \to 0} \;=\;
\frac{1}{2^d \pi^{d/2} \Gamma(d/2)} \,
\int_0^\infty d\kappa \, \kappa^{d-1} \;
\ln\big[ 1 + \frac{\tilde\lambda(\kappa)}{2\kappa}\big] \;.
\end{equation}
Note that the large-$\kappa$ behaviour of the integral is:
\begin{equation}
\int_0^\infty d\kappa \, \kappa^{d-1} \;
\ln\big[ 1 + \frac{\tilde\lambda(\kappa)}{2\kappa}\big]
\, \sim \, \frac{1}{2} \, \int_0^\infty d\kappa \, \kappa^{d-2} \tilde\lambda(\kappa) \;,
\end{equation}
as in the sharp-$\sigma$ case. Assuming a constant $\tilde\lambda$, and introducing an Euclidean cutoff $\Lambda$ for the $\kappa$-integration, we see that:
\begin{equation}
\int_0^\infty d\kappa \, \kappa^{d-1} \;
\ln\big[ 1 + \frac{\tilde\lambda(\kappa)}{2\kappa}\big]
\, \sim \, \frac{1}{2} \,\tilde\lambda_0 \;
\int_0^\Lambda d\kappa \, \kappa^{d-2} \;,
\end{equation}
where the $\sim$ refers to the large-$\kappa$ behaviour of the integrals only.
The UV divergences in the previous expression are then quite immediate to extract: in $d=1$
(two spacetime dimensions) there is a logarithmic divergence:
\begin{equation}
\big[{\mathcal E}_0\big]_{\epsilon \to 0, div} \;=\;
\frac{\tilde\lambda_0}{4 \pi} \, \; \ln(\frac{\Lambda}{\mu}) \;,
\end{equation}
where $\mu$ is a momentum scale. It should be noted that these divergences
may be interpreted in quantum field theoretic terms. Indeed, when
$\tilde\lambda$ is a constant, the model we consider may be regarded as a
free scalar field theory with $\varphi^2$ insertions; the latter coming
from an expansion of the term proportional to $\tilde\lambda$ in the action
(which is local when $\tilde\lambda$ is a constant). For $n$
insertions (i.e., the term of order $\tilde\lambda^n$) in $d+1$ dimensions,
the superficial degree of divergence $\delta$ of a term contributing to
${\mathcal E}_0$ with $n$ insertions of $\varphi^2$ is, by power counting:
\begin{equation}
\delta = (d+1) + n (d-1)- n (d + 1) \;.
\end{equation}
For the $d=1$ case, we see that only the $n=1$ term (linear in
$\tilde\lambda$) diverges and it does so logarithmically, as we have seen
above. In two spatial dimensions, there is only a linear divergence, in the
$n=1$ term, while for $d=3$ there are divergences for $n=1$ (quadratic) and $n=2$
(logarithmic). Therefore, we expect to have one, two and three
renormalization conditions for $d=1$, $d=2$ and $d=3$, respectively.

To make sense of those divergences, rather that dealing with the one particle
irreducible functions corresponding to the operator insertions, we apply
the renormalization program in a more straightforward way, in terms of the vacuum energy,
as follows: subtracting from the integrand in the
expression for ${\mathcal E}_0$ with a constant $\tilde\lambda$ its MacLaurin
expansion in $\tilde\lambda$  up to an order $\delta+1$, where $\delta$ corresponds to the
higher degree of divergence at the given $d$, we have a convergent expression.

The renormalized energy will then have a finite degree polynomial, whose
coefficients must be fixed by imposing suitable renormalization conditions. For example,
in $d=1$:
\begin{equation}
{\mathcal E}_0 \;=\; c_1 \, \tilde\lambda_0
\,+\, \Delta{\mathcal E}_0\;,
\end{equation}
where $c_1$ is a constant (depending logarithmically on the cutoff),
and $\Delta{\mathcal E}_0$ is the finite expression:
\begin{equation}
\Delta {\mathcal E}_0\;=\;
\frac{1}{2 \pi} \,
\int_0^\infty d\kappa \; \Big[
\ln\big( 1 + \frac{\tilde\lambda_0}{2\kappa}\big)
- \frac{\tilde\lambda_0}{2\kappa}\Big]
\;.
\end{equation}
A simple rescaling shows that $\Delta {\mathcal E}_0$ is also linear in
$\tilde\lambda_0$ (although with a cutoff-independent, finite coefficient).
Thus we conclude that the renormalized energy will have the form:
\begin{equation}
\big[{\mathcal E}_0\big]_{d=2} \;=\; C_1 \, \tilde\lambda_0 \;,
\end{equation}
where $C_1$ requires, to be fixed, to know the vacuum energy at some scale.
In $d=2$, on the other hand,
\begin{equation}
{\mathcal E}_0 \;=\; c_1 \, \tilde\lambda_0 \,+\, c_2 \,
\tilde\lambda_0^2 \,+\, \Delta{\mathcal E}_0\;,
\end{equation}
where now,
\begin{equation}
\Delta {\mathcal E}_0\;=\;
\frac{1}{4 \pi} \,
\int_0^\infty d\kappa  \kappa \; \Big[
\ln\big( 1 + \frac{\tilde\lambda_0}{2\kappa}\big)
- \frac{\tilde\lambda_0}{2\kappa}  + \frac{1}{2}
\big( \frac{\tilde\lambda_0}{2\kappa}\big)^2 \Big]
\;,
\end{equation}
which goes like $\tilde\lambda_0^2$; then:
\begin{equation}
\big[{\mathcal E}_0\big]_{ren} \;=\; C_1 \, \tilde\lambda_0 \,+\, C_2 \,
\tilde\lambda_0^2  \;.
\end{equation}

Finally, in $d=3$, a similar procedure yields:
\begin{equation}
\big[{\mathcal E}_0\big]_{ren} \;=\; C_1 \, \tilde\lambda_0 \,+\, C_2 \,
\tilde\lambda_0^2 \,+\, C_3 \tilde\lambda_0^3 \;.
\end{equation}
As already mentioned, the vacuum energy is simply related to the effective action in the presence of $\varphi^2$ operator insertions; hence the
renormalization conditions on the vacuum energy may be related to conditions for the corresponding one particle irreducible functions. From a more
phenomenological point of view, the meaning of the renormalization conditions is best put in a negative way: they summarize the {\em ignorance\/}
one has about the vacuum energy, when it is divergent.
Namely, one can predict the dependence of the vacuum energy with the coupling constant, except for the first few terms in a McLaurin
expansion of the energy in terms of $\tilde\lambda_0$, the $\kappa \to \infty$ part of that coupling constant.
It is perhaps worth emphasizing that in a real situation there must be a cutoff; hence $\tilde\lambda_0 = 0$, and the
divergences above are replaced by finite, cutoff dependent terms, which one does not need to renormalize, and whose precise form
depends on the details on the defect (its profile, for example).

We see that we would need to impose three of those conditions in order to completely fix the
renormalization constants $C_i$. We agree with the results
of~\cite{Graham}, for the case of a sharp defect.

Finally, note that the would be $n=0$ ($\tilde\lambda$-independent) divergences do not
appear because we measure energies with respect to the vacuum in the
absence of mirrors.
\section{Results about the Casimir energy for two mirrors}\label{sec:double}
We shall now consider the evaluation of the Casimir energy, as a function
of the different parameters, for different profiles.
\subsection{Thin mirrors}\label{ssec:thin}
As a first check of the method, we consider firstly the  case
of thin walls ($\epsilon \to 0$) and arbitrary $\lambda$.  This corresponds, physically, to $\epsilon$ much smaller than the other two parameters
with the dimensions of a length, $a$ and $\tilde\lambda^{-1}$.
Later on we shall also impose the condition that $\lambda$ tends to infinity to recover the
well-known Dirichlet case.

The $\epsilon \to 0$ condition can be easily imposed on $\tilde{\mathcal
K}$; indeed, when $\epsilon \to 0$, one has the following expression for
that kernel:
\begin{equation}
\tilde{\mathcal K}^{-1}(z;z') \,\to\, \delta(z-z')
\,-\,\frac{\tilde{\lambda}(\kappa)}{2\kappa \,+\, \tilde{\lambda}(\kappa)}\;,
\end{equation}
and
\begin{equation}
{\mathcal P}(z;z') \,\to\,
\frac{\tilde{\lambda}(\kappa)}{2\kappa}\,e^{- \kappa \, a}\;.
\end{equation}
Then it is immediate to see that, in the same limit,
\begin{equation}
{\mathcal O}(z;z') \,\to\, \Big(\frac{\tilde{\lambda}(\kappa)}{2\kappa \,+\,
\tilde{\lambda}(\kappa)}\Big)^2 \, e^{-2 \kappa a} \;,
\end{equation}
i.e., it becomes independent of $z$ and $z'$.
Thus the Casimir energy density for $\epsilon \to 0$ is given by the
expression:
\begin{equation}
\tilde{\mathcal E}_0(a) \;=\; \frac{1}{2^d \pi^{d/2} \Gamma(d/2)} \,
\int_0^\infty d\kappa \, \kappa^{d-1} \;
\ln \Big[1 \,-\, \big(\frac{\tilde{\lambda}(\kappa)}{2\kappa \,
+\, \tilde{\lambda}(\kappa)}\big)^2 \, e^{-2 \kappa a}\Big] \;,
\end{equation}
which could be evaluated numerically for different interesting functional forms of
$\tilde{\lambda}$, depending on the material considered.

We can obtain exact results using some particular profiles.
The Dirichlet case is obtained by considering a constant $\tilde{\lambda}$
which tends to infinity.  Note that the integral is convergent for any
finite $\tilde{\lambda}$, thus the limit could be taken after evaluating
the integral over momenta.
Nevertheless, taking the $\tilde{\lambda} \to \infty$ limit before  we can
perform the integral exactly, in any number of
dimensions, $d$. For example,
\begin{equation}
\tilde{\mathcal E}_0(a) \;=\; \left\{
\begin{array}{cl}
-\frac{\pi}{24 a} & {\rm for} \, d=1 \\
-\frac{\zeta(3)}{16 \pi a^2} & {\rm for} \, d=2 \\
-\frac{\pi^2}{1440 a^3} & {\rm for} \, d=3
\end{array}
\right.
\end{equation}

There is another profile for $\lambda$ which allows we to find exact
results, albeit it is unphysical regarding the properties of the function
$\tilde\lambda$. However, it serves the purpose of illustrating a property of the
Casimir effect in the Dirichlet case: consider the profile $\tilde\lambda(\kappa) = 2
\alpha \, \kappa$, where $\alpha$ is a constant,
which has the unphysical property of {\em growing with the frequency and momentum}. In
this case we have:
 \begin{eqnarray}
\tilde{\mathcal E}_0(a) &=& \frac{1}{2^d \pi^{d/2} \Gamma(d/2)} \,
\int_0^\infty d\kappa \, \kappa^{d-1} \;
\ln \Big[1 \,-\, \big(\frac{\alpha}{1+\alpha}\big)^2 \, e^{-2 \kappa a}\Big]
\nonumber\\
&=& I(d,\alpha)\, \frac{1}{a^d} \;,
\end{eqnarray}
where
\begin{equation}
I(d,\alpha) \equiv  \frac{1}{2^d \pi^{d/2} \Gamma(d/2) } \,
\int_0^\infty dx \, x^{d-1} \; \ln \Big[1 \,-\,
\big(\frac{\alpha}{1+\alpha}\big)^2 \, e^{-2 x}\Big]\;,
\end{equation}
is a finite number, depending on the constant $\alpha$ and the
dimension, $d$. Note that this kind of profile produces a
dependence of the energy with the distance that is identical to
the Dirichlet case, although with a smaller coefficient, in spite
of the fact that the coupling constant is a function that grows
with $\kappa$. This is a reflection of the fact that, although
$\tilde\lambda$ is not infinite, its particular form introduces
reflection and transmission coefficients that are independent of
$\kappa$, as in the Dirichlet case.

Of course, when $\alpha \to \infty$, we recover the Dirichlet result, which
shows that one can approach it not just from the constant $\tilde\lambda$
case, but also starting from a rather different profile.

If, on the other hand, $\epsilon$ is negligible with
respect to  $a$, but not necessarily in comparison with ${\tilde\lambda}^{-1}$, we may
use  the approximation:
\begin{equation}
|x_d(z) - x_d(z') + a | \,\sim\, a
\end{equation}
in the expressions defining ${\mathcal P}$ and ${\mathcal Q}$.
Then, from the definition of ${\mathcal O}$, we see that:
\begin{equation}
{\mathcal O}(z;z') \sim \Big(\frac{{\tilde\lambda}(\kappa)}{2\kappa}\Big)^2
\,
\Big[\int dz_2 \int dz_3 \, {\tilde{\mathcal K}}^{-1} (z_2;z_3) \Big] \;
\int dz_1 {\tilde{\mathcal K}}^{-1}(z;z_1) \; e^{- 2 \kappa a } \;,
\end{equation}
which is independent of $z'$.

Then, ${\mathcal O}$ has only one non-vanishing eigenvalue, $\vartheta$, for each $\kappa$, as in the $\epsilon \to 0$
case:
\begin{equation}
\vartheta \;=\;  \Big(\frac{\tilde\lambda (\kappa)}{2\kappa}\Big)^2 \,
\Big[\int dz \int dz' {\tilde{\mathcal K}}^{-1}(z;z') \Big]^2 \;
e^{- 2 \kappa a }\;,
\end{equation}
and
\begin{equation}
\tilde{\mathcal E}_0(a) \;=\; \frac{1}{2^d \pi^{d/2} \Gamma(d/2)} \,
\int_0^\infty d\kappa \, \kappa^{d-1} \;
\ln \Big\{1 \,-\, \big(\frac{\tilde{\lambda}(\kappa)}{2\kappa}\big)^2 \,
 \Big[\int dz \int dz' {\tilde{\mathcal K}}^{-1}(z;z') \Big]^2
 \, e^{-2 \kappa a}\Big\} \;.
\end{equation}
Of course, the result depends on $\epsilon$ because of the object
$\int dz \int dz' {\tilde{\mathcal K}}^{-1}(z;z')$. For the case of the piecewise constant defect, for example,
we find:
\begin{eqnarray}
 \int dz \int dz' {\tilde{\mathcal K}}^{-1}(z;z') &=&
\frac{8}{\pi} \sum_{k=0} \frac{1}{(2 k + 1)^2} \,
\frac{ (2 k + 1)^2 \pi^2 + ( 2 \epsilon \kappa)^2}{  (2 k + 1)^2 \pi^2 + ( 2 \epsilon \kappa)^2 + 2 \epsilon {\tilde\lambda}(\kappa)} \nonumber\\
&=& 1 \,-\, \frac{8}{\pi} \sum_{k=0} \frac{1}{(2 k + 1)^2} \,
\frac{ 2 \epsilon {\tilde\lambda}(\kappa)}{  (2 k + 1)^2 \pi^2 + ( 2 \epsilon \kappa)^2 + 2 \epsilon {\tilde\lambda}(\kappa)} \;.
\end{eqnarray}

We conclude the study of the thin wall case by mentioning that it is
possible to study analytically the net-to-leading term in an expansion in powers of
$\epsilon$ for ${\tilde {\mathcal E}}_0(a)$, by extending the method used in
thin-wall approximation. A rather lengthy calculation
shows  that the  first-order term exactly vanishes. Thus, the first
non-trivial correction to ${\tilde {\mathcal E}}_0(a)$ (if present),
can be of the form $C \epsilon^2/a^{d+2}$, in $d$ spatial dimensions.

We also wish to pinpoint to a phenomenon, which appears when
considering the finite size limit in expressions obtained by using
the eigenvalues corresponding to a particular profile. The
$\epsilon \to 0$ and $\tilde\lambda \to \infty$ limits, which we
have studied exactly at the beginning of \ref{ssec:thin}, have
been found by using the explicit form of ${\tilde{\mathcal
K}}^{-1}$. It corresponds, in fact, to the exact solution when the
mirrors are $\delta$-like potentials. That limit, however, cannot
be taken from the finite-$\epsilon$ expressions we have derived
for particular profiles, since their derivation assumes that, for
all $x_d$, $\sigma_\epsilon(x_d) > 0$, in order to map ${\mathbb
R}$ to a finite interval, finding the eigenfunctions in that
finite region. The use of that mapping to find those
eigenfunctions collapses when $\epsilon \to 0$, however, since
then $\sigma_\epsilon(x_d)$ becomes a $\delta$-function.

\subsection{Finite width mirrors}
We now study some general properties of $\tilde{\mathcal E}_0(a)$,
which are independent of any assumption regarding the form of the
defect.

We first note that, denoting by ${\mathcal O}_{ll'} = \langle \psi_l |
{\mathcal O} | \psi_{l'}\rangle$  the matrix elements of the operator
${\mathcal O}$ in the basis of eigenfunctions of ${\mathcal K}$, and a
similar convention for the ${\mathcal P}$ and ${\mathcal Q}$ kernels, we
have:
\begin{equation}\label{eq:oll}
{\mathcal O}_{ll'} \;=\; \frac{1}{\alpha_l} \,
\sum_m {\mathcal P}_{l m} \frac{1}{\alpha_m} {\mathcal Q}_{ml'} \;.
\end{equation}
To study the properties of $\tilde{\mathcal E}_0$ in more detail,
we consider the explicit form of ${\mathcal P}_{l m}$
and ${\mathcal Q}_{l m}$:
\begin{eqnarray}
{\mathcal P}_{l m} &=& \frac{{\tilde\lambda}(\kappa)}{2\kappa} \int dz \int dz' \, \psi_l(z) e^{- \kappa |x_d(z) - x_d(z') - a |} \psi_m(z') \nonumber\\
{\mathcal Q}_{l m} &=& \frac{{\tilde\lambda}(\kappa)}{2\kappa} \int dz \int dz' \, \psi_l(z) e^{- \kappa |x_d(z) - x_d(z') + a |} \psi_m(z') \,=\, {\mathcal P}_{ml}\;.
\end{eqnarray}
Then,
\begin{equation}\label{eq:oll1}
{\mathcal O}_{ll'} \;=\; \frac{1}{\alpha_l} \,
\sum_m \frac{{\mathcal P}_{l m}{\mathcal P}_{l'm}}{\alpha_m} \;.
\end{equation}
An eigenvalue $\vartheta$ corresponding to ${\mathcal O}$ (required in order to evaluate the Casimir energy by applying (\ref{eq:casimir})) is then
determined by $\sum_{l'}{\mathcal O}_{ll'} v_{l'} = \vartheta v_l$.
Using (\ref{eq:oll1}), and taking into account the fact that $\alpha_l > 0$, this may be equivalently written in matrix form as follows:
\begin{equation}\label{eq:gen}
 A \, v \;=\; \vartheta \, g \,v
\end{equation}
where $A$ is a symmetric and positive definite matrix whose elements are given by
\begin{equation}
A_{kl} = \sum_m \frac{{\mathcal P}_{k m}{\mathcal P}_{l m}}{\alpha_m}
\end{equation}
and $g$ is a diagonal matrix: $g  = {\rm diag}\{\alpha_1,\alpha_2, \ldots  \}$.

Note that (\ref{eq:gen}) has the form of a generalized eigenvalue problem for the symmetric matrix $A$ and the diagonal positive matrix $g$. Taking advantage of the fact that $g$ is diagonal, one can show that the eigenvalues of $A$ can be found as the ones of a  matrix $B$ such that $B_{kl} \equiv \alpha_k^{-\frac{1}{2}} A_{kl} \alpha_l^{-\frac{1}{2}}$. This matrix $B$ may be given an even more convenient form:
\begin{equation}
 B \;=\;  C \, C^\dagger \;,
\end{equation}
where
\begin{equation}
C_{kl} \, \equiv \,\frac{{\mathcal P}_{k l}}{\sqrt{\alpha_k \alpha_l}} \;.
\end{equation}
Being symmetric and definite positive,  there are many available numerical algorithms to compute
the eigenvalues of the matrix $B$.
\section{Conclusions}\label{sec:concl}
In this paper we have computed the Casimir energy for a real scalar field
in different backgrounds that describe finite width, semitransparent
mirrors. The properties of the mirrors are described by $ \lambda(x_0-x_0';
{\mathbf x}_\parallel - {\mathbf
x}_\parallel')$ and $\sigma_\epsilon (x_d)$. The former is  a two-point
function that depends on time and on the parallel coordinates, and
represents frequency-dependent transmission and reflection coefficients of
the mirrors, while the latter depends only  on the coordinate normal to the
mirrors,  and describes the spatial dependence of their electromagnetic
properties. Our starting point, a non local effective action for the
quantum scalar field, can be thought as arising from the interaction with
the degrees of freedom of the mirrors.

The functions $\lambda$ and $\sigma_\epsilon$ act as physical regulators
for the divergences of the zero point energy, and the usual `perfect
conductor' limit can be obtained when $\sigma_\epsilon$
becomes a $\delta$-function and $\lambda$ tends to infinity. In this case,
the effect of the mirrors is to impose Dirichlet boundary conditions on the
scalar field.

We have computed explicitly the self-energy of a mirror using smooth and
piecewise constant  profiles $\sigma_\epsilon$. In both cases, the
self-energy is finite  under the assumption that the mirror becomes
transparent at high frequencies.  We have discussed in detail the
dependence of the results with the cutoff frequency $\omega_c$ at which the
mirror becomes transparent,  and with the width $\epsilon$ of the mirror,
showing that the zero point energy vanishes for $\epsilon \omega_c <1$.
Finally, we have also analyzed the UV divergences that arise as
$\epsilon\to 0$, recovering previous results \cite{Graham}.  We hope that
an analogous  computation of local quantities like $<T_{\mu\nu}>$ will, in this
context, be useful to discuss the gravitational effects of the zero
point fluctuations, without having to deal neither with bulk nor surface
divergences.

For the case of two mirrors, we have discussed some general properties of
the interaction energy, for arbitrary functions $ \lambda$ and
$\sigma_\epsilon$. In the thin wall limit, the expression for the energy is
considerably simpler, and we have evaluated it explicitly for some
particular cases, reproducing also the well known results  for perfect mirrors.

Regarding future research, as already mentioned, it would certainly be of
interest to compute also {\em local\/} quantities, like the energy and pressure
densities. Moreover, a realistic model for the interaction between
the quantum field and the degrees of freedom in the mirror would
allow us to derive a nonlocal effective action suitable for a
detailed analysis of the dissipation effects. This realistic model
would necessarily involve the electromagnetic field. Due to gauge
invariance, on general grounds we expect the nonlocal effective
action to be of the form:
\begin{equation}
{\cal S}_{\rm eff} \,=\,\int d^{d+1}x\,\int d^{d+1}x'\,\,
F_{\mu\nu} (x) K^{\mu\nu\rho\sigma}(x ; x') F_{\rho\sigma}(x')\,\,
,
\end{equation}
where the kernel $K^{\mu\nu\rho\sigma}(x ; x')$ encodes the
electromagnetic properties of the mirrors. Here gauge invariance is inherited from the (assumed)
gauge invariant coupling between the material and the gauge field. On the other hand, this is consistent
with  the limiting case of ideal plates, where the boundary conditions are given in terms of $F_{\mu\nu}$ rather than $A_\mu$.
In spite of the fact that the effective action shall involve derivatives of the gauge field, we expect a conveniently adapted version of the method
will make it possible to consider also this case.

The extension of the results of this paper to non planar and/or
non static mirrors would also be of high interest.

 \section*{Acknowledgements}
C.D.F. thanks CONICET, UNCuyo and ANPCyT for financial support. The work of
F.D.M. and F.C.L was supported by UBA, CONICET and ANPCyT.

\end{document}